\documentclass[useAMS,usenatbib,twocolumn]{mn2e}
\usepackage{epsfig}
\usepackage{amsmath}
\usepackage{amssymb}

\begin{document}

\title{A two-dimensional mixing length theory of convective transport}

\author[P. Lesaffre, S. M. Chitre, A. T. Potter and C. A. Tout]
{Pierre Lesaffre$^{1,2}$\thanks{E-mail: pierre.lesaffre@lra.ens.fr},
Shashikumar M. Chitre$^{3,4}$,
Adrian T. Potter$^{4}$
\newauthor
and Christopher A. Tout $^{4}$\\
$^{1}$ LERMA (LRA), UMR 8112, Observatoire de Paris
$^2$ \'Ecole Normale Sup\'erieure, 24 rue Lhomond, 75231 Paris, France \\
$^{3}$ Centre for Basic Sciences, University of Mumbai, India \\
$^{4}$ Institute of Astronomy, The Observatories, Madingley Road, Cambridge CB3 0HA, UK}

\maketitle

\begin{abstract}

The helioseismic observations of the internal rotation profile of
the Sun raise questions about the two-dimensional (2D) nature of the
transport of angular momentum in stars.  Here we
derive a convective prescription for axisymmetric (2D) stellar
evolution models.
We describe the small scale motions by  a spectrum of unstable linear modes in a
Boussinesq fluid. Our saturation prescription makes use of the angular
dependence of the linear dispersion relation to estimate the anisotropy
of convective velocities. We are then able to provide closed form
expressions for the thermal and angular momentum fluxes with only
one free parameter, the mixing length. 

  We illustrate our prescription for slow rotation, to first order
   in the rotation rate. In this limit, the thermodynamical variables
   are spherically symetric, while the angular momentum depends both
   on radius and latitude. We obtain a closed set of equations for
   stellar evolution, with a self-consistent description for the
   transport of angular momentum in convective regions. We derive the
   linear coefficients which link the angular momentum flux to the
   rotation rate ($\Lambda$- effect) and its gradient
   ($\alpha$-effect). We compare our results to former relevant
   numerical work.

\end{abstract}


\section{Introduction}

Computations in stellar evolution have generally been carried in
one-dimensional (1D) frameworks. Since the seminal paper of
\citet{BV58}, mixing length theory (MLT) has proved to be a very
powerful tool to compute the transport of heat in stars, even though
its underlying assumptions are very often regarded as crude in
comparison to the complexity of the usually highly turbulent
convective flows.  Recently, it has also become clear from
helioseismic observations that the rotational profile of the Sun is
intrinsically two-dimensional \citep[2D, see][ for
  example]{S98}. Moreover, the inclusion of rotation in stars has been
shown to be essential in many phases of stellar evolution
\citep{MMI,YL04} but these simulations usually assume solid-body
rotation within convective regions and a self-consistent treatment of
rotation and convection is still lacking. We shall emphasise in this
work that the characteristics of the angular momentum fluxes depend on
the latitude even for spherically symmetric rotating stars. Indeed,
the properties of the turbulent motions should naturally depend on the
angle between the gravitational field and the angular velocity
vector. It is thus clear that a proper treatment of the evolution of
the rotation profile of stars requires a two-dimensional
description. In this work we { lay the basis of} a self-consistent
MLT formulation for 2D-axisymmetric rotating stars which could be
adapted for future 2D stellar evolution computations. { We
  illustrate our precriptions in the case of slow rotation, to first
  order in the rotation rate. In this limit we recover the classical
  1D stellar evolution set of equations for spherically symetric stars
  with an additional 2D equation for angular momentum transport.} We
also provide a spherically averaged version for practical uses in
current 1D stellar evolution codes including slow rotation, which
allows for a self-consistent treatment of angular momentum transport
in convective regions. { We discuss our results in comparison to
  appropriate existing numerical simulations.}

   We start from the stellar fluid dynamical
equations which we average on a smoothing length scale.  Next, we
consider the equations for perturbed quantities, for which we use a
quasi-linear approximation: namely, we assume the perturbed fields are
composed of { a spectrum of } unstable linear modes. Their amplitude is then determined
by a saturation prescription which takes into accounts some of the
non-linearities. We finally proceed to compute the heat and momentum
fluxes which enter our averaged equations, thus closing our
system. The only parameter in our model is the smoothing length scale,
which we identify with the mixing length. { This enables us to construct the stellar evolutionary equations without invoking additional parameters with accompanying assumptions.}

Earlier \citet{G78}\footnote{\citet{gough2012} extended his earlier
  work to higher order in $\Omega$.} and \citet{DS79} developed very
similar ideas but they adopted slightly different saturation
prescriptions.  They modelled the perturbed quantities with a
{\it single} representative { unstable} mode with unspecified parameters
to characterize its orientation whereas here we use the linear dispersion
relation to infer the { \it full spectrum} of the perturbations.
{We therefore predict the anisotropy without extra parameters. An
  advantage of our approach is that it spells out the underlying
  assumptions which can then form the basis to improve the
  prescription.  } \citet{2003MNRAS.340..969O} followed by
\citet{2010MNRAS.407.2451G} derived dynamical equations for the second
order correlations supplemented by closure relations which introduce a
set of additional non-dimensional parameters. Earlier,
\citet{1997ApJ...482..827C} went even further in the hierarchy of
correlations and provided a set of dynamical equations for quantities
up to third order correlation terms with closure relations
parametrized by even more free parameters.
\citet{1993A&A...276...96K} approximated the effects of turbulence by
a viscous stress tensor and computed the effects of rotation on the
momentum fluxes.  Except for \citet{1997ApJ...482..827C}, all these
authors discussed only solid body rotation. We consider here the
dynamics of convective motions in the presence of large-scale fields,
such as non-uniform rotation and incorporate the presence of a shear
in the calculation of turbulent fluxes.

In Section 2 we describe our framework in a Cartesian geometry and we
present and discuss our MLT prescription, comparing it with the works
of \citet{G78} and \citet{DS79}. We apply it to { slowly rotating}
spherical axisymmetric stars in Section 3. We discuss our results and
conclusions in Section 4. In the Appendix, we give the full closed set
of stellar evolution equations in the limit of slow rotation and we
provide their spherically averaged equivalent for 1D stellar
models. 

\section{General framework}

We start with the equations of fluid dynamics subject to a local gravitational
acceleration $\boldsymbol{g}$ with Cartesian components $g_{i}$. The
mass conservation is given by 
\begin{equation}
\partial_{t}\rho+\partial_{i}(\rho v_{i})=0,
\label{eq:continuity}
\end{equation}
where $\rho$ is the mass density, $v_{i}$ the components of the
velocity and $\partial_{t}\equiv\partial/\partial t$ and
$\partial_{i}\equiv\partial/\partial x_{i}$ are
the partial derivatives with respect to time and each of the three
spatial coordinates.  We have used Einstein's summation convention. 

The momentum conservation leads to

\begin{equation}
\partial_{t}(\rho v_{i})+\partial_{j}(\rho v_{i}v_{j}+p\delta_{ij})=\rho g_{i},
\end{equation}
where $p$ is the pressure and we have neglected viscosity. 

The heat transport equation \citep[see for example][equation
49.4]{landau1987} combined with continuity
(equation~\ref{eq:continuity}) becomes

\begin{equation}
T\partial_{t}(\rho S)+T\partial_{j}(v_{j}\rho S)+\partial_{j}F_{j}=q,
\label{eq:heat-original}
\end{equation}
where $S$ is the specific entropy, $q$ the net heat generation rate,
$T$ the temperature and it is assumed there is no dependence of entropy
on the chemical composition. The radiative flux is given by 

\begin{equation}
F_{j}=-\rho c_{p}\chi\partial_{j}T
\end{equation}
where $c_{p}$ is the heat capacity at constant pressure, $\chi$
is the thermal diffusivity (unit length $\times$ velocity). The spatial
derivatives of both $c_{p}$ and $\chi$ are assumed to be negligible.

\subsection{Average equations}

We now define the sliding average of a quantity $y$ at position
$\boldsymbol{x}$ by

\begin{equation}
\langle y \rangle = \frac{1}{\mathcal{V}} \int_{V(\boldsymbol{x})}  y\,\mathrm{d}^{3}x
\end{equation}
where $V(\boldsymbol{x})$ is a small cube of volume $\mathcal{V}$
centred on position $\boldsymbol{x}$.
We want to find a new set of equations for the averaged quantities.
These constitute our stellar model equations. We use as new variables
the volume and mass weighted averages, 
\begin{equation}
\bar{\rho}=\langle\rho\rangle ,
\end{equation}
\begin{equation}
\bar{v}_{i}=\frac{\langle\rho v_{i}\rangle }{\langle\rho\rangle }
\end{equation}
and
\begin{equation}
\bar{e}=\langle\rho e\rangle /\langle\rho\rangle .
\end{equation}
We then define the residuals with respect to these averages by
\begin{equation}
y'=y-\bar{y}
\end{equation}
for any quantity $y$.

We now make use of the Boussinesq approximations that velocities are
small compared to the sound speed, wavelengths are small compared to
the local scale height and $\rho'\simeq0$ except when coupled with the
gravity $g_{i}$  (cf. \citealp{SV60}). { Although
  \cite{1969JAtS...26..448G} has shown these approximations to be not
  suitable for stellar convective regions, and instead the anelastic
  approximation should be used,} we feel they capture the essential
physics while simplifying the derivation.  In particular, these
approximations allow us to work with nearly incompressible equations
and to have $\langle y'\rangle =0$ for most quantities of interest.
With these approximations, the volume averaged continuity is
unchanged,
\begin{equation}
\partial_{t}\bar{\rho}+\partial_{i}(\bar{\rho}\bar{v}_{i}) = 0.
\end{equation}
The momentum equation becomes
\begin{equation}
\partial_{t}(\bar{\rho}\bar{v}_{i})+\partial_{j}(\bar{\rho}\bar{v}_{i}\bar{v}_{j}+\bar{p}\delta_{ij}+{\cal R}_{ij})=\bar{\rho}\bar{g}_{i} ,
\end{equation}
where we discard $\langle\rho' g'\rangle$, because gravity
is slowly varying and we use the convective
momentum flux (more commonly referred to as the Reynolds stress tensor)
\begin{equation}
{\cal R}_{ij}= \bar{\rho}\langle v'_{i} v'_{j}\rangle .
\end{equation}
Finally the average entropy evolution equation is
\begin{equation}
\bar{T}\partial_{t}(\bar{\rho}\bar{S})+\bar{T}\partial_{j}(\bar{\rho}\bar{v_{j}}\bar{S}+\mathcal{F}_{j})+\partial_{j}F_{j}= \bar{q}\label{eq:ave_entropy}
\end{equation}
where, noting that $(\bar T +  T')^{-1} \approx \bar
T^{-1}(1- T'/T)$ we neglect the non-linear terms $\langle q' T'\rangle $ and
$\langle T'\partial_{j} F'_{j}\rangle$, as is necessary
to recover the classical formulation of MLT,
and define the convective heat flux
\begin{equation}
{\cal F}_{j}= \bar{\rho}\langle S' v'_{j}\rangle .
\end{equation}
We drop the over bars in what follows to ease writing and reading
but they should be assumed in the remainder of this section. In order to
get expressions for the convective fluxes, we now turn to the estimation
of the small scales (or $ y'$ ) quantities.

\subsection{The sub-grid model}

In order to make progress, we make assumptions about the linearity
and scale of the perturbations. These are those usually made for a
local mixing length theory. In the linear theory, we will then apply
some kind of closure condition to determine the amplitudes of turbulent
modes.

We { appeal to} the Boussinesq approximation and compute the difference between
the general equations and their averages in order to obtain governing
equations for the $ y'$ quantities.
The continuity equation for the perturbed velocity becomes
\begin{equation}
\partial_{i} v'_{i}=0.\label{eq:small-continuity}
\end{equation}
The momentum equations yield

\begin{equation}
\partial_{t}(\rho v'_{i})+\partial_{j}[\rho v_{i}\delta
  v_{j}+\rho v_{j} v'_{i}+\rho v'_{i} v'_{j}+\delta
  p\delta_{ij}-{\cal R}_{ij}]=\rho'  g_{i},
\end{equation}
where we have discarded $ g'$ because the gravitational field
is produced by mass deep inside the star.

We now resort to a length scale separation hypothesis: averaged quantities
are assumed to be nearly uniform in the local volume $V$ and perturbed
quantities vary over scales that are much smaller than the diameter
of $V$.
Hence the gradient of ${\cal R}_{ij}$,
\begin{equation}
\partial_{j}{\cal R}_{ij}\ll\partial_{j}[\rho v_{i} v'_{j}+\rho
  v_{j} v'_{i}+\rho v'_{i} v'_{j}+ p'\delta_{ij}],
\end{equation}
the gradients of the perturbed quantities.  Mixing-length theories
which make use of this approximation are generally called \emph{local}
mixing length theories. Most MLT used for practical purposes
in stellar evolution are of this type. 

We further neglect the non-linear terms and so set

\begin{equation}
\rho v'_{i} v'_{j}\approx 0
\end{equation}
and retain only the linear approximation. In accord with the Boussinesq
approximation we also neglect the temporal and spatial variation of the average mass
density. Hence, the perturbed quantities are determined by solving
the linear problem for the scales that fit well inside the local volume
$V$ so that
\begin{equation}
\partial_{t} v'_{i}+ v'_{j}\partial_{j}v_{i}+v_{j}\partial_{j} v'_{i}+\frac{1}{\rho} \partial_{i} p'=\frac{\rho'}{\rho}g_{i}\label{eq:small-momentum}
\end{equation}
where we have retained the shear term from the background velocity.
This is necessary for the redistribution of angular momentum.
With the Boussinesq approximation and neglecting the pressure
perturbations with respect to thermal effects, we write the first law
of thermodynamics as
\begin{equation}
 S'=c_{p}(\frac{T'}{T}-\nabla_{\!\rm a}\frac{p'}{p})\simeq c_{p}\frac{T'}{T}\label{eq:entropy_first_principle}
\end{equation}
where $\nabla_{\!\rm a}=(\partial\log T/\partial\log p)_S$ is the
usual adiabatic gradient.  The entropy conservation
equation is then linearised as

\begin{equation}
\partial_{t}\frac{ T'}{T}+{v}_i\partial_i\left(\frac{ T'}{T}\right) +\frac{1}{c_{p}} v'_{i}\partial_{i}S=\chi\partial_{i}\partial_{i}\left(\frac{ T'}{T}\right),\label{eq:small-energy}
\end{equation}
where we have neglected $ q'$, the time-dependence of $c_{p}$ and
the stratification in the thermal diffusion term.  As is customary in
mixing length theories, we have neglected the term
\begin{equation}
\frac{q}{\rho T}\left(\frac{ q'}{q} - \frac{\rho'}{\rho} -
  \frac{ T'}{T}\right).
\end{equation}
We simply note here that this term could become important when strong
nuclear burning takes place within convective regions.

\subsection{Saturation and amplitude of the linear modes}

We restore some of the non-linear effects by adopting a strong
assumption for the saturation of each mode. We denote by $\lambda_{\rm
  m}$ the smoothing length-scale, a typical scale of the smoothing
volume $V$. At the largest scales within this volume, i.e. scales on
the order of or just below the smoothing length scale $\lambda_{\rm
  m}$, we assume that the saturation of a given mode is due solely to
its own shear. Parasitic instabilities, such as Kelvin-Helmholtz
rolls, feed on the shear motions generated by the parent mode. We
assume that eventually they are responsible for its saturation.  {
  We designate $\widetilde{\bmath{v'}}_{\boldsymbol{k}}$ to be the
  complex amplitude of the Fourier mode of the velocity perturbation
  associated with a wave vector $\boldsymbol{k}$ and define the
  amplitude of the velocity perturbation
\begin{equation}
u_{\bmath{k}}=
\sqrt{|\widetilde{v'}_{1\boldsymbol{k}}|^2+|\widetilde{v'}_{2\boldsymbol{k}}|^2+|\widetilde{v'}_{3\boldsymbol{k}}|^2}.
\label{amplitude}
\end{equation}
}
We write schematically the time evolution of this amplitude by 
\begin{equation}
{\dot u_{\boldsymbol{k}}}\approx\sigma_{\boldsymbol{k}}u_{\boldsymbol{k}}-\sigma_{\mathrm{p}}(\boldsymbol{k},u_{\boldsymbol{k}})u_{\boldsymbol{k}}\label{eq:amp}
\end{equation}
where $\sigma_{\boldsymbol{k}}$ is the { real part of the} linear growth rate of the
mode and $\sigma_{\mathrm{p}}(\boldsymbol{k},u_{\boldsymbol{k}})$
is the { real part of the} growth rate of the parasitic mode responsible for its saturation
and we assume that $\sigma_{\mathrm{p}}$ depends only on $\boldsymbol{k}$
and on the amplitude $u_{\boldsymbol{k}}$ of the parent mode. For
example, in the case of Kelvin-Helmholtz rolls, $\sigma_{\mathrm{p}}$
depends on the component of $\boldsymbol{k}$ orthogonal to $\boldsymbol{ \widetilde{v'}_{\bmath{k}}}$
but, because we have assumed that the motions are almost incompressible, $\sigma_{\mathrm{p}}$
is equal to $ku_{\bmath{k}}$ where $k=|\boldsymbol{k}|$ is the wave number. At saturation,
equation~\eqref{eq:amp} allows to write 
\begin{equation}
\sigma_{\boldsymbol{k}}=\sigma_{\mathrm{p}}(\boldsymbol{k},u_{\boldsymbol{k}})
\end{equation}
which determines the mode's amplitude. For example, if Kelvin-Helmholtz
is the dominant parasitic instability, the saturation is reached when
the velocity is of the order of 

\begin{equation}
u_{\boldsymbol{k}}=\sigma_{\boldsymbol{k}}/k .\label{eq:saturation0}
\end{equation}
Note that, provided $\sigma_{\boldsymbol{k}}$ depends on the direction
of the wave vector $\boldsymbol{k}$, this prescription leads to an
anisotropic amplitude of the velocity.

Similar ideas have been used to assess the saturation of other instabilities.
The parasitic instabilities of the magnetorotational instability (MRI)
were described by \citet{GX94} and \citet{LLB09}. Their role for
the saturation of the MRI was examined independently by \citet{L09}
and \citet{P09}. The prescription we use here is very similar to
that used by \citet{P09} and \citet{P10}. \citet{G10} have used a
slight refinement of these prescriptions to predict the saturation
amplitude of the standing accretion shock instability (SASI) in core-collapse
supernovae.
However, these ideas have so far { been concerned with individual unstable modes}.
We propose here to extend this prescription to { a whole spectrum of} modes. At the largest
scale we use the prescription \eqref{eq:saturation0}. But the smallest
scales are likely to feel the non-linear interactions of the scales
immediately above and below, as in the Kolmogorov cascade \citep{K41}. We therefore
use a { power-law} scaling for each direction individually as a
first approximation. We write $k_{\rm m}=\pi/\lambda_{\rm m}$ the minimum
wave number in our smoothing volume. The resulting closure expression
is then
\begin{equation}
u_{\boldsymbol{k}}=\left(\frac{k}{k_{\rm m}}\right)^{-n} \,\frac{1}{k_{\rm m}} \sigma_{\boldsymbol{k'}},
\label{eq:saturation}
\end{equation}
where $\boldsymbol{k'}=\frac{k_{\rm m}}{k}\boldsymbol{k}$ with $n=11/6$ for
Kolmogorov scaling { or $n=21/10$ for Bolgiano-Obukhov scaling
  \citep{B59,O59}. These will likely bracket the actual spectrum index
  \citep[see][]{Rincon}.} The last factor accounts for the anisotropy
of the driving instability and the first for the energy cascade.  Our
closure equation completely determines the { amplitude of all
  Fourier coefficients of the velociy perturbations.  Once the
  velocity amplitude is known, the linear system of equations for the
  perturbations is used to estimate the amplitude of all other
  perturbed variables relative to the velocity. The numerical study of
  \citet{Rincon} has carefully examined the anisotropy and scaling
  of turbulent convection and we plan to validate our approach with
  such numerical studies.}  Note that the fluxes involve integrals of
$u_{\boldsymbol{k}}^{2}\mathrm{d}^{3}k\propto k^{2-2n}dk$
for $k\geqslant k_{\rm m}$, so our results are only weakly sensitive
to the exponent $n$ as long as it is strictly greater than $3/2$,
otherwise these integrals diverge. This means that the fluxes are
dominated by the largest scales just below the mixing length.  In fact
this conflicts with our assumption of scale separation but this is a
common inconsistency of mixing length theories.  \par Others have
avoided such divergent behaviour by considering only a limited number
of modes. For example, in \citeauthor{G78}'s \citeyearpar{G78}
statistical picture eddies of a given shape are randomly formed, grow
and get disrupted whereas we envisage the sub-grid scale motions to be
a collection of saturated unstable modes.  { \citeauthor{G78} has
  an elaborate time-dependent model for the evolution of an eddy,
  whereas our work assumes a steady-state which saves us from
  specifying the initial conditions. For example, he has to assume
  seeds for the eddies to be isotropically distributed.  Further, he
  has to prescribe a space filling factor for the shape of his
  representative eddy. }

From  equation (4.6) of \cite{G78}, his
definition of $\ell=\pi/k_v$ and if we identify his $\bar{w^2}$ with the
square of the magnitude of the radial component of our velocity
$|\widetilde{v'}_{r\boldsymbol{k}}|^2$ and his $\sigma$ to our $\sigma_{\boldsymbol{k}}$, we arrive at
\begin{equation}
|\widetilde{v'}_{r\boldsymbol{k}}|^{2}=\Lambda \pi^2 \sigma_{\boldsymbol{k}}^{2} (k_{h}^{2}/k^{4}),
\end{equation}
where $k_{h}$ is the magnitude of the horizontal part of the
representative wave vector and $\Lambda$ is a calibrateable
dimensionless constant { which incorporates the anisotropy
  parameter as well as the filling factor}.

\citet{DS79} have
also proposed a similar saturation prescription to ours.  They set 
\begin{equation}
\langle {v'}_r^{2}\rangle =\langle\sigma_{\boldsymbol{k}}^{2}\rangle \pi^{2}/k_{r}^{2}
\end{equation}
for a linear combination of a few modes with similar wave vectors.
{ Both these prescriptions slightly differ mathematically from
  ours. However, the fundamental difference lies in the existence of a
  parameter which prescribes the anisotropy of the velocity field in
  both the works of \citet{G78} and \citet{DS79}, whereas our
  prescription links this anisotropy to physics of the underlying
  instability which generates the perturbations.}

\subsection{Computation of the fluxes}
{ We have hitherto discussed how to determine the modulus of all
  the Fourier coefficients of the perturbations. We now summarize how
  to compute the convective fluxes, which depend on the volume average
  of a product of two perturbed quantities:
\begin{equation}
\langle y'z' \rangle = \frac1{\mathcal V} \int_{V}  y'z' \,{\rm d}^3x \mbox{.}
\end{equation}
  
We assume the volume $V$ is a cube of side $\lambda_{\rm m}$, hence $\mathcal{V}=\lambda_{\rm m}^3$. Any
field on this cube can be represented by its Fourier modes with wave
number coordinates as multiples of $k_{\rm m}=\pi/\lambda_{\rm m}$, for instance:
\begin{equation}
y'(\bmath{x})=\sum_{\bmath{k}/k_{\rm m} \in \mathbb{Z}^3} \widetilde{y'}_{\bmath{k}} e^{i\bmath{k.x}} \mbox{.}
\end{equation}
where $\widetilde{y'}$ denotes the Fourier transform of $y'$.
 We use Parseval's theorem to write
\begin{equation}
  \int_{V} y'z'^{*}\, {\rm d}^3x = \lambda_{\rm m}^3
  \sum_{\bmath{k}/k_{\rm m} \in \mathbb{Z}^3} \widetilde{y'}_{\bmath{k}}\widetilde{z'}^*_{\bmath{k}}
\end{equation}
and
\begin{equation}
  \int_{V} y'^{*}z'\, {\rm d}^3x = \lambda_{\rm m}^3
  \sum_{\bmath{k}/k_{\rm m} \in \mathbb{Z}^3} \widetilde{y'}^{*}_{\bmath{k}}\widetilde{z'}_{\bmath{k}} \mbox{.}
\end{equation}
We take the average of the two previous equations and use the fact
that $y'$ and $z'$ are real fields to get
\begin{equation}
  \int_{V} y'z' \,{\rm d}^3x  = \lambda_{\rm m}^3
  \sum_{\bmath{k}/k_{\rm m} \in \mathbb{Z}^3}  |\widetilde{y'}_{\bmath{k}}||\widetilde{z'}_{\bmath{k}}|\cos(\psi(\bmath{k})) 
\end{equation}
where $\psi(\bmath{k})$ is the phase difference between
$\widetilde{y'_{\bmath{k}}}$ and $\widetilde{z'_{\bmath{k}}}$. We estimate this phase
difference from the linear analysis of the corresponding mode.

 Finally, we approximate the sum on all wave vectors by a continuous integral over the
non-dimensional wave-vector $\widetilde{\bmath{k}}=\bmath{k}/k_{\rm m}$:
\begin{equation}
\langle y'z' \rangle =  \int_{|\widetilde{\bmath{k}}|>1}  |\widetilde{y'}_{\widetilde{\bmath{k}}}||\widetilde{z'}_{\widetilde{\bmath{k}}}|\cos(\psi(\widetilde{\bmath{k}})) \,{\rm d}^3\widetilde{k} \mbox{.}
\label{eq:general-flux}
\end{equation}
}

\section{Slowly rotating axisymmetric stars}

{ The local rotation rate $\Omega$ of a star can be compared to two
  rates of interest to construct dimensionless numbers. On the one hand,
  the inverse of the free-fall time scale yields
  $\epsilon_0=\Omega\sqrt{r/g}$, while the buoyancy
  frequency provides another number $\epsilon=\Omega/N$ where $N$ is the magnitude of
  the Brunt-V\"ais\"al\"a frequency. The latter can also be written
  $\epsilon=\Omega\sqrt{H_S/g}$ where $H_S$ is on the order of the entropy
  scale height. Although most stars have $\epsilon_0 \ll 1$, the entropy
  mixing in convective regions can make $H_S$ very large and
  $\epsilon$ is not necessarily close to zero. For example, our Sun
  has $\epsilon_0=7.4\times 10^{-4}$ at the surface, but $\epsilon$ can
  be of order unity at the bottom of the convective region. In the
  following, we derive the stellar evolution equations to first order
  in the parameter $\epsilon$ and assume $\epsilon_0<\epsilon$. We
  will further assume that the tides are weak and that an axisymmetric
  model about the rotation axis can suffice.}

\subsection{Background state}

In such an axisymmetric star, it is natural to use a spherical
coordinate system, with $r$, $\theta$ and $\phi$ as the radius,
co-latitude and azimuth respectively, for the definition of the
background.  { The hydrostatic pressure balance with centrifugal
  acceleration necessarily implies that the deviation from spherical
  symetry in the thermodynamic quantities $p$ and $\rho$ is on the
  order of $\epsilon_0^2$. In the first order we can therefore safely
  assume that the thermal background depends on the radius $r$ only
  and that the gravitational acceleration is radial. Similarly, since
  meridional circulation is the result of second order terms ($\Omega^2$),
  we neglect it.}  Then the averaged velocity profile consists only of
cylindrical rotation so that

\begin{equation*}
\bar{v}_{\phi}=r \Omega(r,\theta) \sin\theta
\end{equation*}
and

\begin{equation*}
\bar{v}_{r}=\bar{v}_{\theta}=0.
\end{equation*}

According to our assumptions the background must be smooth over
the scale of the volume $V$. This requires that the first derivative
of $\Omega$ with respect to the cylindrical radius is zero on the
axis of symmetry.

\subsection{Linear System of equations for the modes}
\label{linear-system}

We develop the perturbation at a position $\boldsymbol{x_{0}}$ in
terms of local Fourier modes in a local Cartesian frame rotating about
the axis of symmetry with angular velocity
$\Omega_{0}=\Omega(\boldsymbol{x_{0}})$.  
The three axes of the frame $\hat{\bmath{x}}$, $\hat{\bmath{y}}$ and $\hat{\bmath{z}}$
are made to coincide with the local spherical coordinate unit vectors,
$\hat{\bmath{r}}$, $\hat{\bmath{\theta}}$ and $\hat{\bmath{\phi}}$ at $\boldsymbol{x_{0}}$. We
consider only one single mode in this subsection. 
Thus
\begin{equation}
 y'_{\boldsymbol{k}} \equiv\Re[\widetilde{y'}_{\boldsymbol{k}}\,\exp(st+ik_{x}x+ik_{y}y+ik_{z}z)],\label{eq:Fourier}
\end{equation}
where $\widetilde{y'}_{\boldsymbol{k}}$ is the complex amplitude
of the mode under consideration, $i=\sqrt{-1}$ , $x$, $y$
and $z$ are the coordinates in this local Cartesian frame
and $k_x$, $k_y$ and $k_z$ are the three components
of the wave vector in this frame. The three coordinates have the physical
dimension of the radius $r.$ In particular, $\mathrm{d}x=\mathrm{d}r$,
$\mathrm{d}y=r\mathrm{_{0}d}\theta$ and $\mathrm{d}z=r_{0}\sin\theta_{0}\mathrm{d}\phi$.
In principle, shear deforms non-axisymmetric perturbations on a time
scale of the order of the local shear time, which we therefore assume
to be long compared with the growth time $\Re(s)^{-1}$ in order to
apply our linear analysis. Although this is true for the Sun now,
it may not necessarily hold for all stars.  It certainly
is not true for Keplerian discs, where the shear rate is comparable
to the rotation rate and to the inverse of the vertical convective
turnover time-scale. We shall drop  the $\boldsymbol{k}$ subscripts
from the complex amplitudes in this section and the next in order
to ease the readability.

The linearised continuity equation~\eqref{eq:small-continuity} leads
to the incompressibility condition

\begin{equation}
k_{x} \widetilde{v'}_{r}+k_{y} \widetilde{v'}_{\theta}+k_{z} \widetilde{v'}_{\phi}=0.\label{eq:div0}
\end{equation}
The momentum equation~\eqref{eq:small-momentum} now includes an additional
Coriolis term $2\boldsymbol{\Omega}_{0}\times\boldsymbol{ v'}$
because the background velocity in the rotating frame is
$\bar{v}_{\phi}=r\sin\theta[\Omega(r,\theta)-\Omega_{0}]$. { We assume
that the apparent gravitational field (including centrifugal acceleration) is vertical and
write $g=g_{r}$, in accordance with our first order expansion in the rotation rate}. The linearised Euler equations become
\begin{equation}
s \widetilde{v'}_{r}-2\Omega_{0}\sin\theta_{0}\, \widetilde{v'}_{\phi}=-\frac{ik_{x}}{\rho} p'-g\frac{\rho'}{\rho},\label{eq:r_euler}
\end{equation}

\begin{equation}
s \widetilde{v'}_{\phi}-2\Omega_{0}\cos\theta_{0}\, \widetilde{v'}_{\phi}=-\frac{ik_{y}}{\rho} p'\label{eq:th_euler}
\end{equation}
and

\begin{equation}
s \widetilde{v'}_{\phi}+2\Omega_{0}M_{r} \widetilde{v'}_{r}+2\Omega_{0}M_{\theta} \widetilde{v'}_{\theta}=-\frac{ik_{z}}{\rho} p',\label{eq:phi_euler}
\end{equation}
where 

\begin{equation}
M_{r}=\frac{1}{2\Omega r\sin\theta_0}\partial_{r}j\label{eq:Mr}
\end{equation}
and
\begin{equation}
M_{\theta}=\frac{1}{2\Omega r^{2}\sin\theta_0}\partial_{\theta_0}j\label{eq:Mt}
\end{equation}
and the specific angular momentum $j$ is 
\begin{equation}
j=r^{2}\Omega \sin^{2}\theta_0.
\end{equation}
Here, $M_{r}$ and $M_{\theta}$ are two dimensionless quantities
proportional to the spherical coordinates of the gradient of specific angular
momentum at our reference point so that
\begin{equation}
\boldsymbol{M}=\frac{1}{2\Omega r\sin\theta_0}\boldsymbol{\nabla}j.
\end{equation}
For uniform rotation, this vector is simply $\boldsymbol{M}=\boldsymbol{\hat{R}}$
where $\boldsymbol{\hat{R}}$ is the cylindrical radius unit vector.

The Boussinesq approximation means that pressure perturbations are
negligible when compared to density and thermal perturbations.  So the
equation of state becomes

\begin{equation}
\frac{\rho'}{\rho} +\Delta \frac{ T'}{T} =0,\label{eq:eos}
\end{equation}
where
\begin{equation}
\Delta=-\left(\frac{\partial\ln\rho}{\partial\ln T}\right)_{p}
\end{equation}
is the compressibility at constant pressure. 

Finally, the entropy equation~\eqref{eq:small-energy} becomes

\begin{equation}
(s+\chi k^{2})\frac{ T'}{T}=-\frac{1}{g\Delta} N^{2} \widetilde{v'}_{r},\label{eq:energy}
\end{equation}
where 
\begin{equation}
k^{2}=k_{x}^{2}+k_{y}^{2}+k_{z}^{2}
\end{equation}
is the square of the modulus of the wave vector and the thermal Brunt-V\"ais\"al\"a
frequency $N^{2}$ is 
\begin{equation}
N^{2}=\frac{g\Delta}{c_{p}} \partial_{r}S
\end{equation}
{ where we neglect the latitudinal thermal gradients, in accordance with our first order expansion
in the rotation frequency.}
In the following, we drop the $0$ subscripts of $\theta$ and $\boldsymbol{M}$
for the sake of tidiness.

\subsection{Dispersion relation and growth rate}

The set of linear equations \eqref{eq:div0} to \eqref{eq:energy} forms an eigenvalue problem for
$s$. { Its dispersion relation is a cubic in $s$ which we express as:

\begin{align}
s^3 +&[1- (\hat{\bmath{k}}. \hat{\bmath{r}})^2] N^2 s \nonumber\\
 +& 2\Omega (\hat{\bmath{k}}. \hat{\bmath{\phi}})  (\hat{\bmath{R}}-\hat{\bmath{M}}).[s^2\hat{\bmath{k}} + (\hat{\bmath{k}}. \hat{\bmath{\theta}}) N^2  \hat{\bmath{\theta}} ]
+ 4\Omega^2 ( \hat{\bmath{k}}.\hat{\bmath{\Omega}})\hat{\bmath{k}}.(\hat{\bmath{M}}\times \hat{\bmath{\phi}}) \, s \nonumber\\
+&  \chi k^2 [ s^2 
+2\Omega  \hat{\bmath{k}}.(\hat{\bmath{R}}-\hat{\bmath{M}}) s 
+2\Omega^2  ( \hat{\bmath{k}}.\hat{\bmath{\Omega}})\hat{\bmath{k}}.(\hat{\bmath{M}}\times \hat{\bmath{\phi}})]=0 
\label{eq:disp}
\end{align}
where $\hat{\bmath{k}}$ and $\hat{\bmath{\Omega}}$ are the unit
vectors along $\bmath{k}$ and $\bmath{\Omega}$. Without thermal
diffusion, this dispersion relation depends only on the direction of
the wave vector $\hat{\bmath{k}}$ and not on its magnitude. For { uniform}
rotation ($\hat{\bmath{M}}=\hat{\bmath{R}}$) and $\chi=0$, we recover
the results from both \cite{1951ApJ...114..272C} and
\cite{DS79}. For axisymmetric modes
($\hat{\bmath{k}}. \hat{\bmath{\phi}}=0$), we recover the dispersion
relation of \cite{1967ApJ...150..571G} without viscosity.

  We now set $\chi=0$ and turn to evaluate the largest real part of
  the roots of the dispersion relation. We will seek the first order
  expansion of the growth rate in the form $s=Ns_0+\Omega
  s_1=N(s_0+\epsilon s_1)$.  For the largest real root at zeroth
  order, we get
\begin{equation}
  s_0=\sqrt{1-(\hat{\bmath{k}}. \hat{\bmath{r}})^2}
\end{equation}
provided $N^2<0$, which is the condition for instability. This
expression shows that the fastest growing modes have zero radial wave
number so that vertical convective plumes are preferred. Our
saturation prescription based on the directional dependence of the
growth rate will be sensitive to this.

 The first order of the largest real root is
\begin{equation}
 s_1= (\hat{\bmath{k}}. \hat{\bmath{\phi}})(\hat{\bmath{R}}- \hat{\bmath{M}})
.[\hat{\bmath{k}}+\frac{\hat{\bmath{k}}. \hat{\bmath{\theta}}}{1-(\hat{\bmath{k}}. \hat{\bmath{r}})^2} \hat{\bmath{\theta}}].
\end{equation}

Close to marginal stability, the marginal root $s_{\rm m}=0$ without
rotation can have the largest real part for slow rotation. However,
$s_{\rm m}$ is first order in $\epsilon$ and the associated fluxes are
of order $\epsilon^{2}$ and we safely neglect it.

Finally, since $s_0$ and $s_1$ are always real, we simply take 
\begin{equation}
  \sigma_{\bmath{k}}=N(s_0+\epsilon s_1)
\label{eq:growth}
\end{equation}
when $N^2<0$ and $\sigma_{\bmath{k}}=0$ otherwise.
}

\subsection{Convective fluxes}

{ Both the first and second order of the growth rate are real
  numbers, consequently the linear system of equations \eqref{eq:div0}
  to \eqref{eq:energy} introduces no phase shift between the perturbed
  fields involved. In our notations, the phase shift which enters the
  expression  \eqref{eq:general-flux} for the flux is
  $\psi(\bmath{k})=0$ or $\pi$ for all pairs of variables of interest.
}

\subsubsection{Kinetic energy}

We start by deriving a useful relation between variables $\widetilde{v'}_{r}$ and $\widetilde{v'}_{\theta}$.
We use equation \eqref{eq:div0} to express the variable $\widetilde{v'}_{\phi}$ in terms of the other two components of the velocity. Then,
we combine equations \eqref{eq:th_euler} and \eqref{eq:phi_euler} to
eliminate the variable $p'$ and so find the relationship between
$\widetilde{v'}_{r}$ and $\widetilde{v'}_{\theta}$ to be

\begin{align}
-[\alpha \beta s - & \beta \gamma 2\Omega M_{r}+\alpha \gamma 2\Omega\cos\theta]\widetilde{v'}_{r} \nonumber \\
=[(1-\alpha^2)s - & \beta \gamma 2\Omega
  M_{\theta} + \beta \gamma 2\Omega\cos\theta]\widetilde{v'}_{\theta}.\label{eq:relationship}
\end{align}
 where we defined the more compact variables
\begin{align}
\alpha=& \hat{\bmath{k}}. \hat{\bmath{r}}=k_x/k ,\\
\beta= & \hat{\bmath{k}}. \hat{\bmath{\theta}}=k_y/k \mbox{ and}\\ 
\gamma=& \hat{\bmath{k}}. \hat{\bmath{\phi}}=k_z/k.
\end{align} Here  $\alpha^{2}+\beta^{2}+\gamma^{2}=1$.

We now develop to first order in $\epsilon$ the ratio of $\widetilde{
v'}_{\theta}$ over $ \widetilde{v'}_{r}$ from
relation \eqref{eq:relationship} to formally obtain

\begin{equation}
 \widetilde{v'}_{\theta}/ \widetilde{v'}_{r}=-\frac{\alpha\beta}{\beta^{2}+\gamma^{2}}+\gamma\epsilon f(\alpha^{2},\beta^{2}),
\end{equation}
where $f$ is a complicated function of $\alpha^{2}$ and $\beta^{2}$
of order 1. The terms involving $s_{1}$ cancel.
We now use equation~\eqref{eq:div0} to get

\begin{equation}
 \widetilde{v'}_{\phi}/ \widetilde{v'}_{r}=-\frac{\alpha\gamma}{\beta^{2}+\gamma^{2}}+\gamma\epsilon f(\alpha^{2},\beta^{2})
\end{equation}
and we apply our saturation prescription \eqref{eq:saturation} to
arrive at

\begin{equation}
\sigma_{\boldsymbol{k'}}^{2}k_{\rm m}^{-2}\tilde{k}^{-2n}= |\widetilde{v'}_{r}|^{2}+ 
|\widetilde{v'}_{\theta}|^{2}+ 
|\widetilde{v'}_{\phi}|^{2}= \widetilde{v'}_{r}^{2}\,(1-4\epsilon\alpha\beta\gamma f ) /(\beta^{2}+\gamma^{2}).
\end{equation}
In the expression \eqref{eq:general-flux} we separate the integral over the magnitude of the wave vector from the integral over all possible directions of the wave vector. We determine that
\begin{equation}
\langle  v_{r}^{'2}\rangle =
\int_{1}^{+\infty}\frac{N^{2}}{k_{\rm m}^{2}}\tilde{k}^{-2n}\,4\pi\tilde{k}^{2}\mathrm{d}\tilde{k}\,
\int_{\alpha^2+\beta^2+\gamma^2=1}s_0^{2}(\beta^{2}+\gamma^{2}) \, \frac{\mathrm{d}\alpha \mathrm{d} \beta \mathrm{d}\gamma}{4 \pi}
\label{eq:flux-vrr}
\end{equation}
to the lowest (zeroth) order in $\epsilon$ and we use $\tilde{k}=k/k_{\rm m}$. The first order in $\epsilon$
is odd in $\gamma$ and its integration over the unit sphere yields a
zero contribution. 
We perform the integral on the non-dimensional
modulus of the wave vector $\tilde{k}$ and write
\begin{equation}
\langle  v_{r}^{'2}\rangle =\frac{4N^{2}\lambda_{\rm m}^{2}}{\pi(2n-3)}\,F_{rr}
\end{equation}
with 
\begin{equation}
F_{rr}=\langle s_0^{4} \rangle_{\rm S}\simeq 0.533
\end{equation}
where $\langle \rangle _{\rm S}$ denotes averaging   over the unit sphere and we used $s_0^2=1-\alpha^2=\beta^2+\gamma^2$. With Kolomogorov scaling ($n=11/6$), 
\begin{equation}
\label{eq:vrr}
\langle  v_{r}^{'2}\rangle =\frac{6}{\pi}\lambda_{\rm m}^{2}N^{2}F_{rr}.
\end{equation}
{   Note that with Bolgiano-Obukhov scaling ($n=21/10$), the prefactor $6/\pi$ decreases to $10/3\pi$ which is about twice smaller. We retain Kolomogorov scaling in the following. }
 The other diagonal
components of the Reynolds-stress tensor are 
\begin{equation}
\label{eq:vtt}
\langle  v_{\theta}^{'2}\rangle =\frac{6}{\pi}\lambda_{\rm m}^{2}N^{2}F_{\theta\theta},
\end{equation}
with
\begin{equation}
F_{\theta\theta}=\langle \alpha^2\beta^2\rangle _{\rm S}
\end{equation}
and
\begin{equation}
\label{eq:vpp}
\langle  v_{\phi}^{'2}\rangle =\frac{6}{\pi}\lambda_{\rm m}^{2}N^{2}F_{\phi\phi},
\end{equation}
with
\begin{equation}
F_{\phi\phi}=\langle \alpha^2\gamma^2\rangle _{\rm S}=F_{\theta\theta}\simeq0.067.
\end{equation}
Our model predicts { a strong} anisotropic distribution of velocities
with motions mostly in the radial direction. 
In accordance with the symetry of the problem, our model predicts equipartition
 between the azimuthal and latitudinal directions.

{ 

\cite{K04} compute these quantities in a number of simulations of
convection including rotation. Our small parameter $\epsilon=\Omega/N$
translates in their notations as
$\epsilon=\frac12(\frac{PrTa}{Ra})^\frac{1}{2}$. Their simulation with
  $Co=1$ corresponds to our $\epsilon=0.09$. On the other hand we need
  the thermal diffusion timescale to be small before the rotation
  timescale because we neglected thermal diffusion (and viscosity),
  and we require $\chi/H_p^2/\Omega$ to be small where $H_p$ is the pressure scale height. The value of this
  parameter is $0.17$ for their simulation with $Co=1$ and bigger for
  lower rotation rates, so we consider only their results at
  $Co=1$. Using $\lambda_{\rm m}=\frac12H_p$ in equations \eqref{eq:vrr},
  \eqref{eq:vtt} and \eqref{eq:vpp} we find 
$\langle\bmath{{v'}^2}
  \rangle^\frac{1}{2}=0.120$ which is only slightly bigger than the
  value $0.090$ which they find (viscous damping or a smaller
  $\lambda_{\rm m}$ could bring these values closer to one another).
  We also predict the ratio of horizontal to vertical motions 
$  \langle v'_{\theta} v'_{\theta} + v'_{\phi} v'_{\phi} \rangle / \langle v'_{r}
  v'_{r} \rangle = 0.125 $ instead of their value of 0.186.  Our
  saturation prescription probably overestimates the anisotropy
  because it neglects the tendency to isotropy at small scales down
  the turbulent cascade \citep[cf.][]{Rincon}.  Nevertheless, our model correctly accounts
  for the fact that the anisotropy does not depend on the latitude.  }

\subsubsection{Thermal fluxes}

From equation~\eqref{eq:energy}, we write 

\begin{equation}
\langle \frac{T'}{T} v'_{i}\rangle =\frac{6}{\pi}\frac{\lambda_{\rm m}^{2}N^{3}}{g}F_{Ti}\label{eq:dvdT},
\end{equation}
with 
\begin{equation}
F_{Tr}=\langle s_0^{3}\rangle _{\rm S}\simeq0.589
\end{equation}
and 

\begin{equation}
F_{T\theta}=-\langle s_0\alpha\beta\rangle _{\rm S}=0
\end{equation}
to the lowest order in $\epsilon$.  {  The latitudinal thermal flux is zero to first order, consistent with the direction of the thermal gradients being vertical.} 

The average equation for the evolution of thermal energy is 
\begin{equation}
T \partial_{t}S+\frac{T}{r^{2}}\partial_{r}(r^{2}{\cal F}_{r})-\boldsymbol{\nabla.}(\rho c_{p}\chi\boldsymbol{\nabla}T)=q,
\end{equation}
with the thermal convective flux given by

\begin{equation}
{\cal F}_{r}=\frac{1}{2}\rho c_{p}\int_{0}^{\pi}d\theta\,\sin\theta\langle  v'_{r}\frac{T'}{T}\rangle .
\end{equation}

Using equation~\eqref{eq:dvdT} we write 
\begin{equation}
{\cal F}_{r}=\rho c_{p}\,\frac{6}{\pi}\, F_{Tr}\,\frac{\lambda_{\rm m}^{2}N^{3}}{g},
\end{equation}
which we further develop into the more familiar thermal diffusive
flux
\begin{equation}
{\cal F}_{r}=-D_{T}\,\rho c_{p}\,\Delta \partial_{r}S,
\end{equation}
with the effective diffusion coefficient 
\begin{equation}
D_{T}=\frac{6}{\pi}\, F_{Tr}\, N\lambda_{\rm m}^2\simeq0.68 N\lambda_{\rm m}^2
\end{equation}
  MLTs traditionally make use of a diffusion coefficient of the form
\begin{equation}
D_{\rm MLT}=\frac13 N \ell_{\rm mix}^2
\end{equation}
{ where $\ell_{mix}$ is the mixing length.}
Comparing this expressions to the usual 1D MLT, we can readily
identify our smoothing length $\lambda_{\rm m}$ with the mixing length
to a numerical factor of order one.

{ In the simulations of \cite{K04} with $Co=1$, they compute the
  eddy heat conductivity, $\chi_{rr}=\langle v'_r T'/T \rangle g/N^2$
  in our notations. They compute the ratio $\chi_{rr}/\nu_t$ where
  $\nu_t=\langle {v'}^2  \rangle^\frac12d/3$ and $d$ is the
  size of the convective zone (see their figure 19) and find it is
  between 0.5 and 0.6 depending on the latitude. With $\lambda_{\rm m}=\frac12 H_p$, 
we predict a slightly
  bigger value of 0.67 for this number, which is overestimated by about
  the same factor than for the r.m.s. velocity (a lower value for
  $\lambda_{\rm m}$ would fix both numbers at the same time).}

\subsubsection{Momentum fluxes}

  The expressions for { momentum} fluxes are a bit more
  complicated. It has become common practice (see
  \citealp{1989drsc.book.....R}) to separate the momentum fluxes into
  a term linear in the rotation frequency (the $\Lambda$-effect) and a
  term which depends on the gradients of the rotation frequency
  ($\alpha$-effect), rather than the gradients of specific angular
  momentum which we have used here. Therefore we offset the quantities
  $M_{r}$ and $M_{\theta}$ by their respective values for solid body
  rotation to compare more directly with previous work. {  Each momentum
  flux develops into a linear combination of terms, characterized by four
  constant coefficients.} For instance, we develop the radial momentum flux as
\begin{equation}
\frac{\langle  v'_{r} v'_{\phi}\rangle
}{\frac{6}{\pi}\lambda_{\rm m}^{2}N\Omega}
= F_{r\phi, Mr}(M_{r}-\sin\theta)
+ F_{r\phi, M\theta}(M_{\theta}-\cos\theta)
+ F_{r\phi, \cos\theta}\cos\theta
+ F_{r\phi, \sin\theta}\sin\theta.
\end{equation}
Since $ F_{r\phi, M\theta}=F_{r\phi, \cos\theta}=0$, this reduces to 
\begin{equation}
\label{eq:radial-momentum-flux}
{\langle  v'_{r} v'_{\phi}\rangle
}
= {\frac{6}{\pi}\lambda_{\rm m}^{2}N\Omega} \, (\frac12 F_{r\phi, Mr}\frac{\partial \ln \Omega}{\partial \ln r}
+ F_{r\phi, \sin\theta})\,\sin\theta
\end{equation}

with
\begin{equation}
F_{r\phi, Mr} = \langle -2s_0^3(\alpha^2+\beta^2)\rangle_{\rm S}\simeq -0.687,
\end{equation}
and
\begin{equation}
F_{r\phi, \sin\theta} = \langle -2s_0\beta^2\rangle _{\rm S}=\langle -s_0^3\rangle _{\rm S}=-F_{Tr}\simeq -0.589.
\end{equation}

The two terms in expression \eqref{eq:radial-momentum-flux} { represent respectively} the
$\alpha$-effect and the $\Lambda$-effect (see
\citealp{1989drsc.book.....R}).  { The radial $\alpha$-effect is
  linked to differential rotation and is diffusive in character. This was also
  found by \citet{HG01} at the equator ($\theta=\pi/2$), but some
  quantities in their expression are defined only implicitly which
  makes a direct comparison difficult.}  In the case of solid body
rotation, we predict a $\Lambda$-effect in the form
\begin{equation}
\Lambda_{r\phi}=\frac{6}{\pi} \, F_{r\phi, \sin\theta}   \,\lambda_{\rm m}^{2}N\Omega \sin\theta\simeq-1.12\lambda_{\rm m}^{2}N\Omega \sin\theta.
\end{equation}
This compares very well with the work of both
\citet{1993A&A...276...96K} and \citet{2010MNRAS.407.2451G} in the
slow rotation limit (see in particular equations~(75) and~(76) of
\citealp{2010MNRAS.407.2451G}). However, we note the $\Lambda$-effect
of \citet{1993A&A...276...96K} is essentially due to density gradients
which we have neglected here, { and they find a $\Lambda$-effect
  with the opposite sign compared to us}.  Note that
\citet{2010MNRAS.407.2451G} present their results as a function of
anisotropy but, as they point out, the anisotropy is not arbitrary in
their framework {as in ours}.  In effect, the numerical coefficient in front of
their $\Lambda$-term depends on the values of their closure parameters
$C_{1}$, $C_{2}$, $C_{6}$ and~$C_{7}$ and their expression does not
differ from that of \citet{1993A&A...276...96K}, except possibly for
the numerical value of the pre-factor { and the sign which could
  be either positive or negative. } Although the notion of a $\Lambda$-term was not used at
that time, both \citet{G78} and \citet{DS79} have such a term in their
formulation and correctly estimate its form in the slow rotation
limit.

\begin{figure}
\includegraphics[width=10cm]{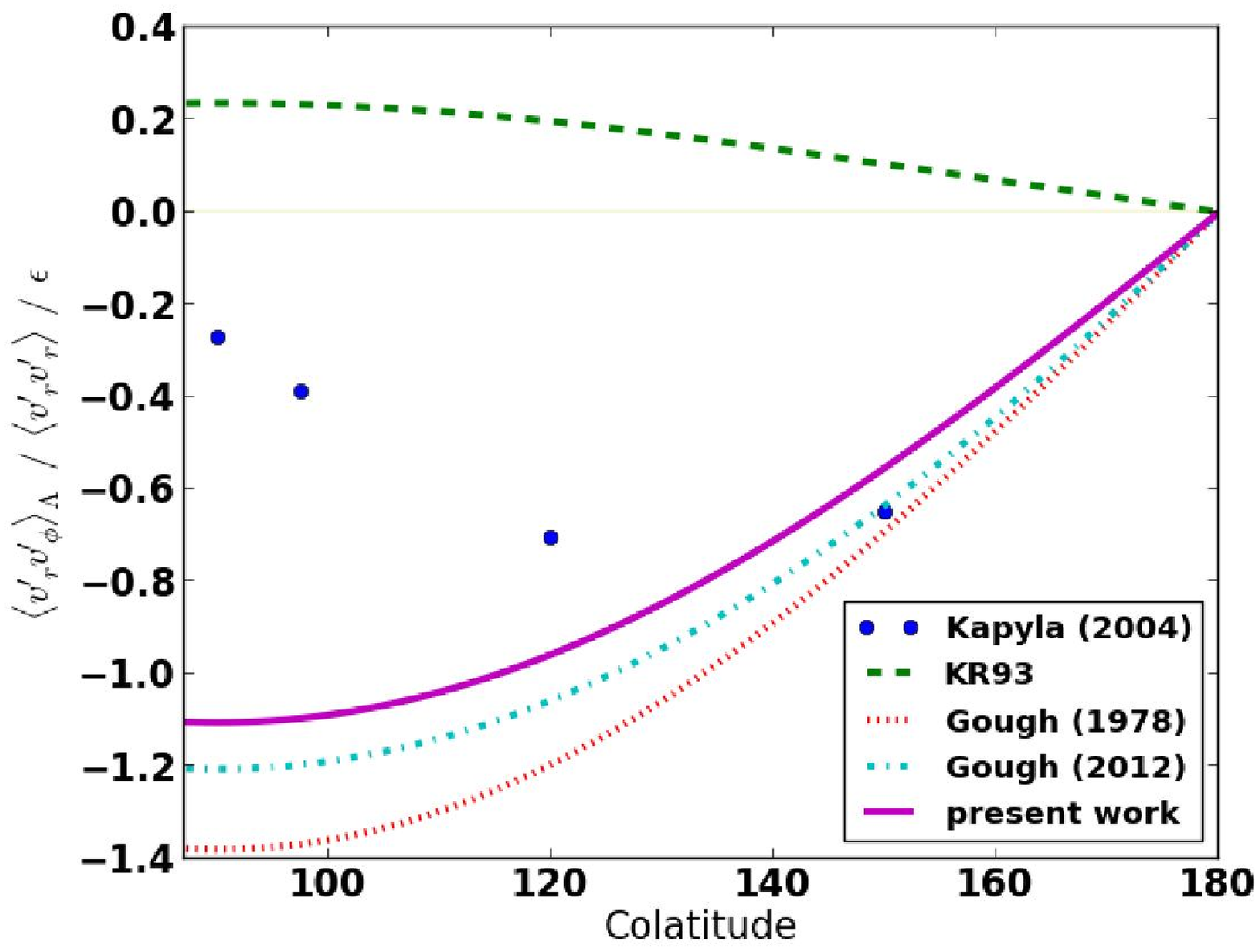}
\caption{Normalised radial angular momentum flux due to the
  $\Lambda$-effect, $\langle v'_{r} v'_{\phi}\rangle_\Lambda /\langle
  v'_{r} v'_{r}\rangle / \epsilon$, according to simulations in
  \citet{K04} (blue dots), to the predictions of
  \citet{1993A&A...276...96K} (green), to both predictions by
  \citet{G78} (red, first order in $\epsilon$), \citet{gough2012}
  (cyan) and to our predictions (magenta).  The numbers from the
  numerical simulations are corrected from the shear (see table 3 in
  \citealp{K04}). To plot the results by \citet{1993A&A...276...96K},
  we use the effective viscosity $\nu_t$ defined as $\nu_t=\langle
  {v'}^2 \rangle^\frac12d/3$ where $d$ is the size of the convective
  region in their numerical setup and we took $\langle v'_{r}
  v'_{r}\rangle$ as measured in the simulations by \citet{K04}. To
  plot the results by Gough, we used the parameter $\Phi=1.9$ as
  measured from the simulations.
 \label{comparison}}
\end{figure}

{ Simulations of \citet{Chan}, \citet{Rieutord} and \citet{K04} all
  find a negative $\Lambda$-effect for slow rotation, in agreement
  with our results. However, we over-estimate by a large amount (up to
  a factor 4 near the equator) the value of the transport coefficient
  compared to the simulations of \citet{K04}, as seen in figure
  \ref{comparison}. The numbers extracted from the simulations are
  corrected from large scale shear flows which appear in their
  simulations. We use the numbers from their table 3, which shows the
  corrections themselves are of the same order as the measured radial
  momentum fluxes.  }

In a similar way we write the latitudinal momentum flux as
\begin{equation}
\label{eq:latitudinal-momentum-flux}
{\langle  v'_{\theta} v'_{\phi}\rangle }
      ={\frac{6}{\pi}\lambda_{\rm m}^{2}N\Omega} \, \frac12
 F_{\theta\phi,
       M\theta}\frac{\partial \ln \Omega}{\partial \theta} \sin\theta
\end{equation}
with
\begin{equation}
F_{\theta\phi, M\theta}=\langle \frac{2}{s_0^3}\alpha^{2}\beta^{2} (\gamma^4-\alpha^2\beta^2-\beta^4)\rangle _{\rm S}\simeq -0.123.
\end{equation}

 {The latitudinal $\alpha$-effect is also diffusive but with a diffusion
  coefficient about six times smaller than the radial one.} The latitudinal $\Lambda$-effect for solid body rotation and our
 vertical entropy gradient is absent { to first order in
   $\Omega$}. This is in agreement with all of \citet{G78},
 \citet{DS79}, \citet{1993A&A...276...96K} and
 \citet{2010MNRAS.407.2451G}, though in the case of \citet{DS79} the
 latitudinal-azimuthal balance of kinetic energy is needed to cancel
 this term.  { This is also consistent with the
   results of \cite{K04} who find this term is much smaller than its
   radial counterpart in the limit of slow rotation.} 

{ The average equation for the transport of angular momentum can be
  found in \cite{D85} (equation (3)). To first order the meridional
  circulation is absent and with our notations the angular momentum
  transport equation may be written as
\begin{equation}
\rho\partial_{t}  r^{2} \sin^{2}\theta\,\Omega+\frac{\sin\theta}{r^{2}} \partial_{r}(r^{3}\,{\cal R}_{r\phi} )+\frac{1}{\sin\theta} \partial_{\theta}(\sin^{2}\theta\,{\cal R}_{\phi \theta} )=0
\label{eq:ave-momentum-1}
\end{equation}

}

We now consider the special case of spherical symmetry which is more
useful for 1D stellar evolution codes. For this purpose we take
$\Omega$ to be a function of $r$ only, so
$M_{\theta}=\cos\theta$ { and the latitudinal transport of momentum vanishes.
It is however customary} to  integrate equation~\eqref{eq:ave-momentum-1} over the angles
$\frac{1}{r}\int_{0}^{\pi}d\theta\,\sin\theta\,\times...$ in order
to eliminate the $\partial_{\theta}$ term so that

\begin{equation}
\rho
\frac{4}{3}r^{2}\partial_{t}\Omega +\frac{1}{r^{2}}\partial_{r}(r^{3}\bar{\cal
  R}_{r\phi})=0,
\end{equation}
with
\begin{equation}
\bar{\cal R}_{r\phi}=\int_{0}^{\pi}d\theta\,\sin^{2}\theta \rho \langle v'_{r} v'_{\phi}\rangle 
\end{equation}
which also reads
\begin{equation}
\bar{\cal R}_{r\phi}=\frac{6}{\pi}\rho\lambda_{\rm m}^{2}N\Omega\,\int_{0}^{\pi}d\theta\,\sin^{2}\theta\,\left[M_{r}\, F_{r\phi, Mr}+\sin\theta\,(F_{r\phi, \sin\theta}-F_{r\phi, Mr})\right].
\end{equation}
We put the last expression back in the average momentum equation to obtain
\begin{equation}
\partial_{t}r^{2}\Omega+\frac{1}{\rho r^{2}}\partial_{r}\left\{ \rho r^{2}\,\frac{1}
{\pi}F_{r\phi, Mr}\,\lambda_{\rm m}^{2}N \left[\partial_{r} (r^{2}\Omega)+
\left(\frac{F_{r\phi, \sin\theta}}{F_{r\phi, Mr}}-1\right)r\Omega\right]\right\} = 0,
\end{equation}
which shows that specific angular momentum is diffused with a diffusion
coefficient 
\begin{equation}
D_{r^{2}\Omega}=-\frac{1}{\pi}F_{r\phi, Mr}\lambda_{\rm m}^{2}N\simeq0.19\, D_{\mathrm{T}}.
\end{equation}

The $\Lambda$-effect
yields an advection term which can be combined with the specific angular momentum
gradient to provide
\begin{equation}
\partial_{t}r^{2}\Omega-\frac{1}{\rho r^{2}}\partial_{r}\left\{ \rho r^{2}\, D_{r^{2}\Omega}\, r^{2}\Omega \partial_{r}\ln\left[r^{1+\frac{F_{r\phi, \sin\theta}(0)}{F_{r\phi, Mr}(0)}}\Omega\right]\right\}=0
\end{equation}
which, after we evaluate the exponent of $r$ in the logarithm, predicts a steady
rotational profile in $\Omega\propto r^{-1.86}$.  Thus the
$\Lambda$-effect offsets the constant specific angular momentum
profile ($\Omega\propto r^{-2}$) by only a small amount.  This
contrasts with most 1D studies of stellar rotation which assume solid
body rotation in convection zones \citep[e.g.][]{MMI,H00}.  \citet{P11}
studied the effects of varying the specific angular momentum
distribution in 1D stellar models and found that the change in the
total angular momentum and additional shear generated at the boundary
between convective and radiative regions can have a significant effect
on the evolution of a star.

\section{Conclusion}
Using a generalized mixing length prescription, we have derived a
self-consistent set of equations for axisymmetric 2D stellar evolution
which includes a description of convective transport of angular
momentum and heat. In the appendix A we list the full set of equations
required to model the evolution of 2D stellar interiors { at first
  order in $\Omega/N$} as well as their 1D spherically averaged
equivalents.

The thermal and momentum fluxes in radial and latitudinal directions
are linked to the properties of the most unstable local linear modes.
In this respect our work in essence follows the spirit that
\citet{G78} pioneered to estimate the fluxes due to small scale
turbulent motions.  However, our approach uses the angular directional
dependence of the convective linear growth rate and determines the
orientation of the convective motions. Thus, our prescription uses
only one parameter, the smoothing length $\lambda_{\rm m}$, which is
readily seen to correspond to the mixing length in the 1D limit.  We
have also studied the dynamics of convective motions in the presence
of an arbitrary rotation field, with radial and latitudinal shear, as
well as a radial and latitudinal thermal stratification.  We provide
simplified expressions relevant for special cases which can readily be
incorporated in stellar evolution codes when the rotation is slow {
  to first order in $\Omega/ N$.  The second order immediately brings
  features such as meridional circulation, non radial effective
  gravity and thermal gradients, and all terms of the dispersion
  relation need to be retained. In the future, we hope to be able
  incorporate these ingredients in our formalism as well as to include
  magnetic fields.}

\section*{Acknowledgements}

{ We should like to express our grateful thanks to the referee for
  producing a thorough and constructive report of the paper and making
  suggestions which have greatly improved its presentation.}  We thank
Douglas Gough for reading an earlier version of our manuscript and for
bringing his pioneering work to our attention.  We also thank Steve
Balbus, Fran\c{c}ois Rincon and Michel Rieutord for stimulating
discussions.  PL gratefully acknowledges support from the French
embassy in the UK while he benefited from an Overseas Fellowship at
Churchill College when this work began in year 2009.  PL also
acknowledges financial support from "Programme National de Physique
Stellaire" (PNPS) of CNRS/INSU, France. CAT also thanks Churchill
College for his Fellowship while SMC enjoyed the use of College's
accommodation while supported by the IOA's STFC visitors' grant and AP
thanks the STFC for his studentship.

\appendix

\section[]{Summary of stellar evolution equations for slow rotation}
\begin{table}
\begin{center}
\begin{tabular}{ccc}
\hline 
Coefficient & Expression & Value\tabularnewline
\hline 
\hline 
$F_{rr}$ & $\langle  s_0^{4}\rangle _{\rm S}$ &  0.533 \tabularnewline
$F_{\theta\theta}=F_{\phi\phi}$ & $\langle  \alpha^{2}\beta^{2}\rangle _{\rm S}$  & 0.067 \tabularnewline
\hline 
\hline 
$F_{Tr}$ & $\langle  s_0^{3}\rangle _{\rm S}$ &  0.589 \tabularnewline
\hline 
\hline 
$F_{r\phi, Mr}$ & $\langle  -2s_0^3(\alpha^2+\beta^2) \rangle _{\rm S}$ & -0.687 \tabularnewline
$F_{r\phi, \sin\theta}=-F_{Tr}$ &
$\langle -s_0^3\rangle _{\rm S}$
 &  -0.589 \tabularnewline
\hline
$F_{\theta\phi, M\theta}$
 &
$\langle \frac{2}{s_0^3}\alpha^{2}\beta^{2} (\gamma^4-\alpha^2\beta^2-\beta^4)\rangle _{\rm S}$
 &  -0.123 \tabularnewline
\hline 
\end{tabular}
\end{center}

\caption{{ Coefficients} relevant to the various correlations involved in the fluxes.\label{Table:expressions}
}
\end{table}

We reproduce here equations for the evolution of spherically symmetric
slowly rotating stellar interiors, valid at first order in the rotation rate:
\begin{equation}
\frac{1}{\rho}\partial_{r}p+g=0\label{eq:hydrostat_r-1}
\end{equation}
and
\begin{equation}
T \partial_{t}S+\frac{T}{r^{2}}\partial_{r}(r^{2}D_{T}\rho c_{p}\Delta
\partial_{r}S)-\boldsymbol{\nabla.}(\rho
c_{p}\chi\boldsymbol{\nabla}T)=q
\end{equation}
{ with
\begin{equation}
D_{T}=\frac{6}{\pi}\, F_{Tr}\, N\lambda_{\rm m}^2
\end{equation}
where $N$, the absolute magnitude of the square root of 
\begin{equation}
N^{2}=\frac{g\Delta}{c_{p}}\,\partial_rS 
\end{equation}
 is the buoyancy frequency and $\lambda_{\rm m}$ is our only
 parameter. We suggest to take the smoothing length $\lambda_{\rm m}$
 as a given fraction of the pressure scale height as is usually done
 for the mixing length. Our comparison with numerical simulations and
 classical MLT suggests $\lambda_{\rm m}=\frac12 H_p$ might be a reasonable
 choice}.

Poisson's equation reduces to 
\begin{equation}
g(r)=\frac{G}{r^{2}}\int_{0}^{r}\mathbf{\mathrm{d}}r'\,4\pi\rho r'^{2}
\end{equation}
and the usual boundary conditions are employed. 
The angular momentum evolution follows the equation
\begin{equation}
\rho\partial_{t}  r^{2} \sin^{2}\theta\,\Omega+\frac{\sin\theta}{r^{2}} \partial_{r}(r^{3}\,{\cal R}_{r\phi} )+\frac{1}{\sin\theta} \partial_{\theta}(\sin^{2}\theta\,{\cal R}_{\phi \theta} )=0
\end{equation}

with
\begin{equation}
{{\cal R}_{r\phi} }=
{\frac{6}{\pi}\rho\lambda_{\rm m}^{2}N\Omega} \,
 (\frac12 F_{r\phi, Mr}\frac{\partial \ln \Omega}{\partial \ln r}
+ F_{r\phi, \sin\theta})\,\sin\theta
\end{equation}
and
\begin{equation}
{{\cal R}_{\phi \theta} }=
{\frac{6}{\pi}\rho\lambda_{\rm m}^{2}N\Omega} \, \frac12
 F_{\theta\phi,
       M\theta}\frac{\partial \ln \Omega}{\partial \theta} \sin\theta \mbox{.}
\end{equation}
 We summarize in table~\ref{Table:expressions} the linear coefficients
 $F_{j}$ needed to determine the convective fluxes.

 When the rotation rate $\Omega$ is taken spherically symetric, we obtain
\begin{equation}
\partial_{t}r^{2}\Omega-\frac{1}{\rho r^{2}}\partial_{r}\left\{ \rho
r^{2}  0.19 D_{T}\, r^{2}\Omega
\partial_{r}\ln\left[r^{1.86}\Omega\right]\right\} = 0
\end{equation}
  for the transport of angular momentum.

\end{document}